# Are Baby Boomers' Non-Work Trip-Making Behavior Different than Millennials? Lessons Learned from NHTS Data[†]

Latif Patwary[1*]; Md Sami Hasnine[2]; and Majbah Uddin[3]

[1]National Transportation Research Center, Oak Ridge National Laboratory, 1 Bethel Valley Rd, Oak Ridge, TN 37830; Email: patwarya@ornl.gov
[2]Department of Civil and Environmental Engineering, Virginia Tech, Blacksburg, VA 24061; Email: hasnine@vt.edu
[3]National Transportation Research Center, Oak Ridge National Laboratory, 1 Bethel Valley Rd, Oak Ridge, TN 37830; Email: uddinm@ornl.gov
[*]Corresponding Author

## ABSTRACT

This paper presents a comparison between Millennials' and Baby Boomers' non-work travel behaviors using data from the 2017 National Household Travel Survey. Bootstrapped segmented ordered logit models are employed to capture the variability in travel preferences and trip frequency across these generational groups, providing more robust insights into their non-work travel. Millennials, particularly those who work from home, are found to have a negative association with higher non-work trip frequency, whereas Baby Boomers have a positive association with higher non-work trip frequency. The model results show that female Millennials who are heads of households are more likely to make non-work trips but less likely when living in urban areas. Ride sharing among Baby Boomers shows a higher association with non-work travel compared to Millennials. These insights could have implications for travel demand management, as shifting travel patterns necessitate adjustments in infrastructure investments and management strategies to support effective long-term transportation planning.

## INTRODUCTION

Traditional four-step models and even simplified activity-based models primarily rely on peak demand to understand commuters' behavior. The rationale is that if the commuter's behavior during peak hours can be understood well, a similar method could be easily transferrable to the other types of trips. There has been a huge shift in commuting patterns (Salon et al. 2022; Moos, Pfeiffer, and Vinodrai 2017). In particular, many workers are telecommuting instead of traveling during peak demand hours. Research also shows that telecommuting individuals are more likely to make increased non-work trips. Non-work travel, including trips for shopping, recreation, social activities, and errands, is a significant component of overall travel demand (Ma, Liu, and Chai 2015). Unlike work-related travel, which is often predictable and concentrated in specific time frames, non-work travel is more flexible and diverse and occurs throughout the day. With the rise of work-from-home trends, non-work travel is becoming an increasingly dominant aspect of travel behavior. The travel patterns of Millennials and Baby Boomers have evolved in distinctly different ways, reflecting broader societal and technological changes. Various studies also show that Baby Boomers' (birth year between 1946 and 1964) travel behavior is different from that of Millennials (birth year between 1981 and 1996). However, it is not clear how Baby Boomers' non-work trip-making behavior is different from that of Millennials. Therefore, it is critical to investigate if Baby Boomers' travel behavior differs from that of Millennials and what factors affect this behavioral shift. Understanding generational differences is essential for shaping transportation planning, as both cohorts influence infrastructure and mobility service needs in distinct ways.

[†] This manuscript has been authored by UT-Battelle, LLC, under contract DE-AC05-00OR22725 with the US Department of Energy (DOE). The US government retains and the publisher, by accepting the article for publication, acknowledges that the US government retains a nonexclusive, paid-up, irrevocable, worldwide license to publish or reproduce the published form of this manuscript, or allow others to do so, for US government purposes. DOE will provide public access to these results of federally sponsored research in accordance with the DOE Public Access Plan (http://energy.gov/downloads/doe-public-access-plan).

To date, only a handful of studies looked at the travel behavior of Millennials and Baby Boomers. A European study analyzed travel and mobility choice differences between Baby Boomers and Millennials using descriptive statistics and logistic regression models (Colli 2020). The study found that Baby Boomers are more car dependent than the Millennials. This study found that Millennials are moving towards more sustainable transportation choices. Also, Zhang and Li (2022) found that Millennials consistently drive fewer miles daily than Baby Boomers. Polzin et al. (2014) only investigated the Millennials' travel behavior and choices of living arrangements. This study found that since most Millennials live with their parents, there is no urgency in getting driving licenses. This delays in getting their driving license, forcing Millennials to rely on sustainable mode choices. Jamal and Newbold (2020) investigated the generational travel behavior of Millennials and older adults based on their literature synthesis. They argued that urban and rural landscapes also impact how Millennials and older adults make their travel choices, and this choice-making behavior is much more complex to address using simple models. The study also found that older adults are less likely to shift towards innovative and sustainable mobility solutions. Ho and Loo (2020) studied multiple generations, classifying them as pre-war, Baby Boomers, and Millennials. Unlike other studies, this study looked at the mobility challenges faced by the pre-war generations due to health issues. Baby Boomers and Millennials have much more flexible travel choices than the pre-war generations. Despite a few studies looking at generational travel behavior, there is a need for the empirical-based analysis between Baby Boomers and Millennials. A few studies only performed a scoping review of the existing literature, whereas other studies either adopted descriptive analysis or linear models. Previous studies did not investigate whether similar socio-technological factors impact Baby Boomers and Millennials differently. To this end, this study uses a large-scale household-level travel survey and estimates advanced econometric models to understand the factors affecting the non-work travel behavior of Baby Boomers and Millennials.

## METHODS

### Data

This study utilizes data from the 2017 National Household Travel Survey (NHTS). The NHTS collects comprehensive daily travel data, linking trip characteristics, such as frequency, distance, duration, mode of transportation, and purpose with personal, household, and vehicle attributes. The focus of this research is on two generational cohorts: Millennials and Baby Boomers, identified using the age variable ($R\_AGE\_IMP$) from NHTS data. Millennials are defined as individuals aged 22-36 years, while Baby Boomers are those aged 53-71 years at the time of the NHTS survey. The trip and household files were merged with person-level data to create a combined dataset for analysis. Data processing included the removal of outliers and entries with incomplete or ambiguous responses, such as "appropriate skip," "prefer not to answer," "not ascertained," and "missing." After these adjustments, the final dataset consists of 30,581 Millennials and 24,281 Baby Boomers nationally.

The dependent variable is non-work trip frequency, an ordinal variable capturing the number of trips taken for purposes such as shopping, recreation, social activities, and errands. The variable is categorized into four categories, i.e., no trips, 1-4 trips, 5-8 trips, and more than 8 trips. The frequency distribution in Table 1 indicates that the majority of both Millennials and Baby Boomers engage in 1-4 non-work trips, with relatively similar patterns observed across the groups.

The explanatory variables included capture personal, household, and location-specific characteristics, providing a comprehensive view of the factors influencing non-work travel behavior. Gender variable shows an equal distribution of males and females in both cohorts (51% male vs. 49% female). Driving status, another key factor, indicates that most respondents are licensed drivers, with 98% of Millennials and 96% of Baby Boomers reporting that they can drive. Education level highlights notable differences between the two groups. Millennials exhibit lower rates of higher education attainment, with 48% holding a bachelor's or graduate degree, compared to 59% of Baby Boomers. Lifecycle stages are distinct across the two groups. Millennials are predominantly adults without children (83%), reflecting their younger age and lifestyle preferences, while Baby Boomers are more evenly split, with 54% living without children and 42% having children under 15 years old. Household income also varies significantly. Baby Boomers have a higher proportion of lower-income households (31% earning less than $50,000 annually) compared to Millennials (24%).

Work-related variables reveal similarities between the two cohorts, with 85% of Millennials and 88% of Baby Boomers working full-time. Despite the growing trend of remote work in recent years, working from home in both cohorts was low in 2017. The distance between home and work is calculated considering the great circle distance, which measures the shortest path between two points on earth's sphere. Most respondents in both groups (93% of Millennials and 95% of Baby Boomers) live within 30 miles of their workplace. The flex-time variable is included to account for whether individuals have flexible working hours, a factor that can significantly influence travel behavior. 46% of Millennials reported having flexible working hours, compared to only 38% of Baby Boomers. All these differences in sociodemographic and travel attributes warrant the need for further investigation into the non-work of Millennials and Baby Boomers.

**Table 1. Descriptive Statistics**

| Variables | Description | Millennials (N= 30,581) | | Baby Boomers (N= 24,281) | |
|---|---|---|---|---|---|
| | | Frequency | Percent | Frequency | Percent |
| Non-work trip frequency (Dependent Variable) | No trips | 7,551 | 25% | 6,329 | 26% |
| | 1-4 trips | 16,968 | 55% | 13,554 | 56% |
| | 5-8 trips | 5,218 | 17% | 3,791 | 16% |
| | >8 trips | 844 | 3% | 607 | 3% |
| Head of household | No | 10,015 | 33% | 12,044 | 50% |
| | Yes | 20,566 | 67% | 12,237 | 50% |
| Gender | Male | 15,535 | 51% | 12,433 | 51% |
| | Female | 15,046 | 49% | 11,848 | 49% |
| Driving status | No | 489 | 2% | 937 | 4% |
| | Yes | 30,092 | 98% | 23,344 | 96% |
| Education | High school or less | 6,331 | 21% | 3,561 | 15% |
| | Some college | 9,684 | 32% | 6,512 | 27% |
| | Bachelor | 7,266 | 24% | 8,498 | 35% |
| | Graduate | 7,300 | 24% | 5,710 | 24% |
| Lifecycle | Adults without children | 20,320 | 83% | 11,326 | 54% |
| | Adults with youngest child 0-15 years | 2,059 | 8% | 8,899 | 42% |
| | Adults with youngest child 16-21 years | 2,188 | 9% | 928 | 4% |
| Full-time worker | No | 4,559 | 15% | 2,848 | 12% |
| | Yes | 26,022 | 85% | 21,433 | 88% |
| Flex-time | No | 16,557 | 54% | 15,045 | 62% |
| | Yes | 14,024 | 46% | 9,236 | 38% |
| Work from home | No | 30,575 | 100% | 24,277 | 100% |
| | Yes | 6 | 0% | 4 | 0% |
| Distance to work | >30 miles | 2,107 | 7% | 1,291 | 5% |
| | ≤ 30 miles | 28,474 | 93% | 22,990 | 95% |
| Weekend | No | 24,498 | 80% | 19,352 | 80% |
| | Yes | 6,083 | 20% | 4,929 | 20% |
| Race | White | 26,281 | 86% | 19,272 | 79% |
| | Black | 1,978 | 6% | 1,685 | 7% |
| | Others | 2,322 | 8% | 3,324 | 14% |
| Household income | <$50,000 | 7,217 | 24% | 7,411 | 31% |
| | $50,000 to $99,999 | 10,695 | 35% | 9,027 | 37% |
| | $100,000 to $149,999 | 6,993 | 23% | 4,907 | 20% |
| | ≥$150,000 | 5,676 | 19% | 2,936 | 12% |
| Home ownership | No | 3,886 | 13% | 9,748 | 40% |
| | Yes | 26,695 | 87% | 14,533 | 60% |
| Car sharing | No | 30,450 | 100% | 24,030 | 99% |
| | Yes | 131 | 0% | 251 | 1% |
| Ride sharing | No | 29,094 | 95% | 19,430 | 80% |
| | Yes | 1,487 | 5% | 4,851 | 20% |
| Urban | No | 7,360 | 24% | 3,781 | 16% |
| | Yes | 23,221 | 76% | 20,500 | 84% |

**Bootstrapped Ordered Logit Models**

The bootstrap method is a resampling technique that generates multiple datasets by randomly sampling with replacement from the original data. Introduced by Efron (1992), bootstrapping is widely used in statistical modeling to improve the robustness and reliability of parameter estimates, particularly in cases of small or imbalanced samples. This approach allows for the generation of empirical distributions for estimated parameters, facilitating statistical inference without relying on strict distributional assumptions (Chernick 2011). In other words, Bootstrapping is particularly useful when dealing with skewed distributions or rare events among independent variables. In our dataset, distributions of several variables are skewed. For example, work from home show a highly skewed distribution, with only a few respondents indicating that they work from home. This imbalance can lead to unreliable parameter estimates and underrepresentation of these cases in standard models. By applying bootstrapping, we replicate cases of rare events to increase their representation in the model, which helps stabilize parameter estimates and enhances model reliability. Support for this approach can be found in the literature. For example, Davison and Hinkley (1997) highlight bootstrapping as a robust tool for addressing issues related to non-normal or skewed distributions in data, particularly when the sample size for specific subgroups is limited. Furthermore, studies, such as Mouratidis et al. (2019), in travel behavior research have used bootstrapping to improve model performance in cases of skewed independent variable distributions. Alternative modeling approaches, such as multinomial logit or generalized ordered logit models, could be considered, but they could introduce additional complexity in interpreting results while offering limited advantages in this context.

The dependent variable, $Y_i$, represents the frequency of non-work trips and is modeled as an ordinal variable with four categories ($k = 0,1,2,3$). The ordered logit model assumes an underlying latent variable $Y_i^*$ such that:

$$Y_i^* = X_i \beta + \epsilon_i \,,$$

Where $X_i$ is a vector of explanatory variables for individual $i$, $\beta$ is a vector of coefficients to be estimated, and $\epsilon_i$ is the error term that follows a logistic distribution.

The observed dependent variable $Y_i$ is related to the latent variable $Y_i^*$ through threshold values $\tau_k$, where:

$$Y_i = \begin{cases} 0 & if\ Y_i^* \leq \tau_0, \\ 1 & if\ \tau_0 < Y_i^* \leq \tau_1, \\ 2 & if\ \tau_1 < Y_i^* \leq \tau_2, \\ 3 & if\ Y_i^* > \tau_2 \end{cases}$$

The probability of observing each category $k$ is given by:

$$f(x) = \begin{cases} F(\tau_0 - X_i \beta), & if\ k = 0, \\ F(\tau_k - X_i \beta) - F(\tau_{k-1} - X_i \beta), & if\ 1 \leq k \leq K-1, \\ 1 - F(\tau_{K-1} - X_i \beta), & if\ k = K \end{cases}$$

where $F(.)$ is the cumulative logistic distribution function.

In the bootstrapped framework, the model is estimated repeatedly across the 1,000 resampled datasets (Mouratidis, Ettema, and Næss 2019), and the final parameter estimates are averaged over all replications. The variance of the estimates is computed using the empirical distribution generated by the bootstrap replications, which allows for robust confidence interval estimation.

**Model Segmentation**

Figure 1 shows the framework of the study, providing a summary of the data, variables, and methods used. As indicated, we further employed a segmented modeling approach for estimating how non-work travel is different between Millennials and Baby Boomers based on different attributes. Specifically, we estimated separate models for each group's non-work trips. Initially, a pooled model incorporating data from both cohorts was developed. Subsequently, drawing on the methodologies of Misra & Atkins (2018) and Patwary & Khattak (2024), we conducted segmented modeling for Millennials and Baby Boomers to capture the unique characteristics influencing each group's travel patterns.

***Test:*** To determine whether the segmented models significantly differ from the pooled model, we employed a chi-square ($\chi^2$) test, calculated as:

$$-2\big[LL(pooled\ model) - LL\big(Millennials_{only}\big) - LL\big(Baby\ Boomers_{only}\big)\big] \sim \chi^2_{df}$$

Where, $degree\ of\ freedom\ (df) = K_1 + K_2 - K$ , where $K_1$ and $K_2$ represent the degrees of freedom associated with the segmented models for Millennials and Baby Boomers, respectively, while $K$ denotes the degrees of freedom for the pooled model.

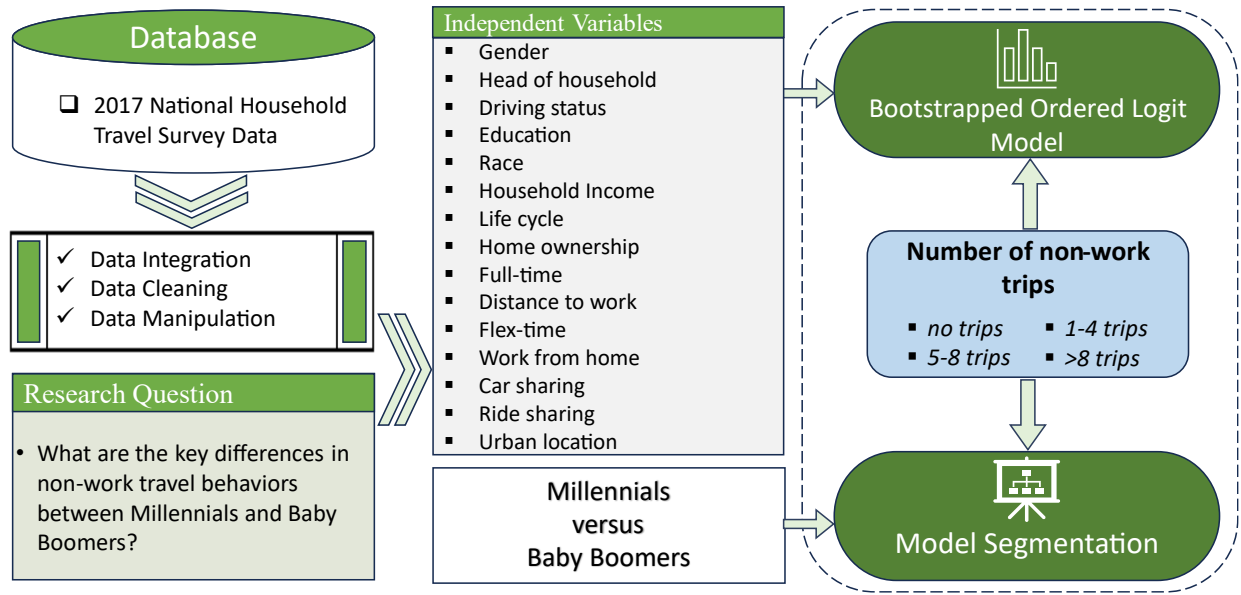

**Figure 1. Study framework**

**RESULTS AND DISCUSSION**

Table 2 and Table 3 show the segmented model results, the former one shows the factors affecting non-work travel for Millennials while the latter shows for Baby Boomers. Model significance tests show both models are statistically significant. Note that the results of the pooled model are not included here. By using the following $\chi^2$ test, we test the null hypothesis that the pooled model is similar to the segmented models:

$$-2\big[LL(pooled\ model) - LL\big(Millennials_{only}\big) - LL\big(Baby\ Boomers_{only}\big)\big] \sim 159.84_{df=27}$$

The resulting *p*-value is close to zero, allowing us to reject the null hypothesis with 99% confidence. This indicates that the segmented models offer a substantially better explanation of the data compared to the pooled model. In other words, we found statistically significant differences between Millennials and Baby Boomers in their non-work travel behavior. Therefore, we can discuss the results based on the differences in contributing factors that influence the non-work travel behavior between Millennials-only and Baby Boomers-only models.

In addition to Tables 2 and 3, Figure 2 illustrates the marginal effects of select key variables influencing non-work trip frequency among Millennials and Baby Boomers. There are several key differences we found from the analysis. The *head of the household* variable is found to be statistically significant and positive. However, it is only significant in the Babby Boomers model. Being the head of the household is associated with an increase in the non-work travel for Baby Boomers. This may indicate that Baby Boomers, as heads of households, often have more responsibilities, such as managing errands and family-related activities, leading to higher non-work travel. *Gender* is significant and positive in both cohorts. Specifically, compared to male Baby Boomers, female Baby Boomers are 1.4% and 3.7% more likely to make "1-4 trips" and "5-8 trips", respectively, for non-work purposes. Whereas these percentages for the same categories are lower for female Millennials. Furthermore, the interaction between *gender* and *head of the household* reveals additional insights into the non-work travel behavior of the generations. The interaction is positive and statistically significant for Millennials-only model. This suggests that female Millennials who are heads of households are even more likely to engage in non-work travel compared to their male counterparts. This aligns with findings that younger women are balancing both professional and household responsibilities, which increases their travel demands (Cao and Mokhtarian 2005). However, for Baby Boomers, this interaction effect is negative but not statistically significant.

Both cohorts show a significant positive relationship between *urban* residence and non-work travel, indicating that living in urban areas increases the likelihood of making non-work trips. This is consistent with the availability of diverse opportunities and destinations within close proximity in urban settings (Mouratidis, Ettema, and Næss 2019). Another important observation is that Millennials living in *Urban* areas tend to make more non-work trips than the ones living in the rural areas. This is similar to the results found for Baby Boomers. Interestingly, the interaction between *gender* and *urban* provides additional insights. Specifically, female Millennials who live in urban areas are less likely to make non-work trips. This is the opposite from the results we found for Baby Boomers, who are more likely to make such travel. *Race* shows similar associations across the two cohorts. Individuals from racial groups classified as "others" are significantly less likely to make non-work trips compared to Whites, as indicated by the negative coefficient.

Another key attributable difference we found in the household *lifecycle* variable. It shows a significant influence on non-work travel for both cohorts. Millennials with children aged 0–15 years are more likely to make non-work trips compared to those without children, as these trips may often involve errands and childcare-related activities. This association is found to be even higher among Baby Boomers. Interestingly, Baby Boomers with children aged 16–21 years are less likely to make non-work trips, possibly due to the increased independence of older children.

*Full-time* work negatively impacts non-work trip frequency for both Millennials and Baby Boomers, as indicated by the significant negative coefficients. This result is expected, as full-time workers typically usually have less time available for discretionary travel during weekdays. Furthermore, *flex-time* is positively associated with non-work travel for both cohorts. The marginal effects suggest that individuals with flexible work schedules are more likely to engage in non-work trips, as flexible hours may allow greater freedom to run errands or pursue leisure activities during non-peak hours. This finding aligns with previous research that highlights the role of work schedule flexibility in shaping travel behavior (Wöhner 2022). Meanwhile, the effect of *working from home* differs between Millennials and Baby Boomers. Baby Boomers working from home have positive association with non-work travel. Specifically, compared to the Baby Boomers who don't work from home, Baby Boomers working from home are 8.6% and 2.1% more likely to make "5-8 trips" and ">8 trips" for non-work purposes, respectively. For Millennials, although it is marginally not-significant (*p-value=0.13*), work-from-home status is negatively associated with non-work trips. It suggests that remote work by Millennials may reduce travel needs. However, this finding for Millennials should be interpreted with caution as it is not statistically significant.

*Car sharing* is found to be not significant for either cohort. However, *ride sharing* is significant and positively associated with non-work trips for both cohorts, indicating that access to ridesharing services facilitates discretionary travel. Baby Boomers exhibit a slightly higher association with rideshare use, potentially reflecting greater reliance on such services as an alternative to personal driving in older generations.

### Table 2. Results for the Millennials-only Model

| Variable (Dependent Variable: Non-work trips) | Coef. | Bootstrap std. err. | *p*-value | Marginal Effect | | | |
|---|---|---|---|---|---|---|---|
| | | | | No trips | 1-4 trips | 5-8 trips | >8 trips |
| Head of household (base: no), yes | 0.044 | 0.038 | 0.254 | -0.024 | 0.003 | 0.017 | 0.004 |
| Gender (base: male), female | 0.109 | 0.055 | 0.047 | -0.025 | 0.003 | 0.018 | 0.004 |
| Driving status (base: no), yes | 0.453 | 0.088 | 0.000 | -0.088 | 0.029 | 0.049 | 0.010 |
| Education (base: high school or less) | | | | | | | |
|     Some college | 0.206 | 0.033 | 0.000 | -0.039 | 0.012 | 0.023 | 0.005 |
|     Bachelor | 0.416 | 0.034 | 0.000 | -0.076 | 0.016 | 0.049 | 0.010 |
|     Graduate | 0.416 | 0.035 | 0.000 | -0.076 | 0.016 | 0.049 | 0.010 |
| Lifecycle (base: adults without children) | | | | | | | |
|     Adults with youngest child 0-15 years | 0.199 | 0.048 | 0.000 | -0.034 | 0.003 | 0.025 | 0.006 |
|     Adults with youngest child 16-21 years | 0.033 | 0.043 | 0.441 | -0.006 | 0.001 | 0.004 | 0.001 |
| Full-time worker (base: no), yes | -0.458 | 0.034 | 0.000 | 0.075 | -0.002 | -0.060 | -0.014 |
| Flex-time (base: no), yes | 0.287 | 0.023 | 0.000 | -0.051 | 0.008 | 0.035 | 0.008 |
| Work from home (base: no), yes | -0.822 | 0.668 | 0.130 | 0.169 | -0.074 | -0.080 | -0.015 |
| GCD to work (base: >30 miles), ≤ 30 miles | 0.027 | 0.043 | 0.523 | -0.005 | 0.001 | 0.003 | 0.001 |
| Weekend (base: no), yes | 1.165 | 0.027 | 0.000 | -0.171 | -0.039 | 0.169 | 0.040 |
| Race (base: white) | | | | | | | |
|     black | 0.047 | 0.046 | 0.312 | -0.008 | 0.001 | 0.006 | 0.001 |
|     others | -0.245 | 0.043 | 0.000 | 0.046 | -0.012 | -0.028 | -0.006 |
| Household income (base: <$50,000) | | | | | | | |
|     $50,000 to $99,999 | -0.013 | 0.032 | 0.690 | 0.002 | 0.000 | -0.002 | 0.000 |
|     $100,000 to $149,999 | -0.006 | 0.037 | 0.860 | 0.001 | 0.000 | -0.001 | 0.000 |
|     ≥$150,000 | 0.042 | 0.039 | 0.288 | -0.007 | 0.001 | 0.005 | 0.001 |
| Home ownership (base: no), yes | -0.103 | 0.038 | 0.006 | 0.018 | -0.002 | -0.013 | -0.003 |
| Car sharing (base: no), yes | 0.051 | 0.178 | 0.774 | -0.009 | 0.001 | 0.006 | 0.001 |
| Ride sharing (base: no), yes | 0.182 | 0.051 | 0.000 | -0.031 | 0.003 | 0.023 | 0.005 |
| Urban (base: no), yes | 0.157 | 0.035 | 0.000 | -0.018 | 0.004 | 0.012 | 0.002 |
| Interaction 1: Gender * Urban | | | | | | | |

| | | | |
|---|---|---|---|
| (female, yes) | -0.117 | 0.051 | 0.022 |
| Interaction 2: Gender * Head of household | | | |
| (female, yes) | 0.184 | 0.051 | 0.000 |

| **Model Fit Statistics** | |
|---|---|
| Number of observations | 30,581 |
| Wald chi2(24) | 2,634.31 |
| Model significance test | 0 |
| AIC | 63,046.33 |
| BIC | 63,271.19 |
| Log-likelihood | -31,496.20 |
| /cut1 | -0.35401 | 0.105237 |
| /cut2 | 2.342994 | 0.105814 |
| /cut3 | 4.576799 | 0.110231 |

**Table 3. Results for the Baby Boomers-only Model**

| Variable (Dependent Variable: Non-work trips) | Coef. | Bootstrap std. err. | *p*-value | Marginal Effect | | | |
|---|---|---|---|---|---|---|---|
| | | | | No trips | 1-4 trips | 5-8 trips | >8 trips |
| Head of household (base: no), yes | 0.188 | 0.036 | 0.000 | -0.034 | 0.008 | 0.021 | 0.004 |
| Gender (base: male), female | 0.321 | 0.069 | 0.000 | -0.059 | 0.014 | 0.037 | 0.008 |
| Driving status (base: no), yes | 0.444 | 0.068 | 0.000 | -0.088 | 0.033 | 0.046 | 0.009 |
| Education (base: high school or less) | | | | | | | |
| Some college | 0.166 | 0.043 | 0.000 | -0.033 | 0.012 | 0.017 | 0.003 |
| Bachelor | 0.330 | 0.046 | 0.000 | -0.063 | 0.020 | 0.036 | 0.007 |
| Graduate | 0.435 | 0.047 | 0.000 | -0.081 | 0.022 | 0.049 | 0.010 |
| Lifecycle (base: adults without children) | | | | | | | |
| Adults with youngest child 0-15 years | 0.236 | 0.028 | 0.000 | -0.042 | 0.009 | 0.028 | 0.006 |
| Adults with youngest child 16-21 years | -0.166 | 0.068 | 0.015 | 0.032 | -0.011 | -0.018 | -0.003 |
| Full-time worker (base: no), yes | -0.281 | 0.044 | 0.000 | 0.049 | -0.007 | -0.034 | -0.007 |
| Flex-time (base: no), yes | 0.370 | 0.028 | 0.000 | -0.067 | 0.014 | 0.044 | 0.009 |
| Work from home (base: no), yes | 0.649 | 0.254 | 0.011 | -0.101 | -0.006 | 0.086 | 0.021 |
| GCD to work (base: >30 miles), ≤ 30 miles | 0.069 | 0.060 | 0.253 | -0.013 | 0.003 | 0.008 | 0.002 |
| Weekend (base: no), yes | 1.057 | 0.032 | 0.000 | -0.165 | -0.012 | 0.144 | 0.032 |
| Race (base: white) | | | | | | | |
| black | 0.010 | 0.050 | 0.839 | -0.002 | 0.000 | 0.001 | 0.000 |
| others | -0.141 | 0.036 | 0.000 | 0.026 | -0.007 | -0.016 | -0.003 |
| Household income (base: <$50,000) | | | | | | | |
| $50,000 to $99,999 | -0.019 | 0.031 | 0.551 | 0.003 | -0.001 | -0.002 | 0.000 |
| $100,000 to $149,999 | -0.001 | 0.042 | 0.987 | 0.000 | 0.000 | 0.000 | 0.000 |
| ≥$150,000 | -0.042 | 0.048 | 0.390 | 0.008 | -0.002 | -0.005 | -0.001 |
| Home ownership (base: no), yes | -0.086 | 0.028 | 0.003 | 0.016 | -0.004 | -0.010 | -0.002 |
| Car sharing (base: no), yes | 0.132 | 0.114 | 0.250 | -0.023 | 0.004 | 0.016 | 0.003 |
| Ride sharing (base: no), yes | 0.203 | 0.033 | 0.000 | -0.036 | 0.007 | 0.024 | 0.005 |
| Urban (base: no), yes | 0.122 | 0.048 | 0.012 | -0.023 | 0.006 | 0.014 | 0.003 |
| Interaction 1: Gender * Urban (female, yes) | 0.010 | 0.069 | 0.888 | | | | |
| Interaction 2: Gender * Head of household (female, yes) | -0.011 | 0.053 | 0.840 | | | | |

| **Model Fit Statistics** | | |
|---|---|---|
| Number of observations | | 24,281 |
| Wald chi2(24) | | 2,036.67 |
| Model significance test | | 0 |
| AIC | | 49,329.24 |
| BIC | | 49,547.87 |
| Log-likelihood | | -24,637.62 |
| /cut1 | 0.137 | 0.104 |
| /cut2 | 2.875 | 0.107 |
| /cut3 | 5.096 | 0.114 |

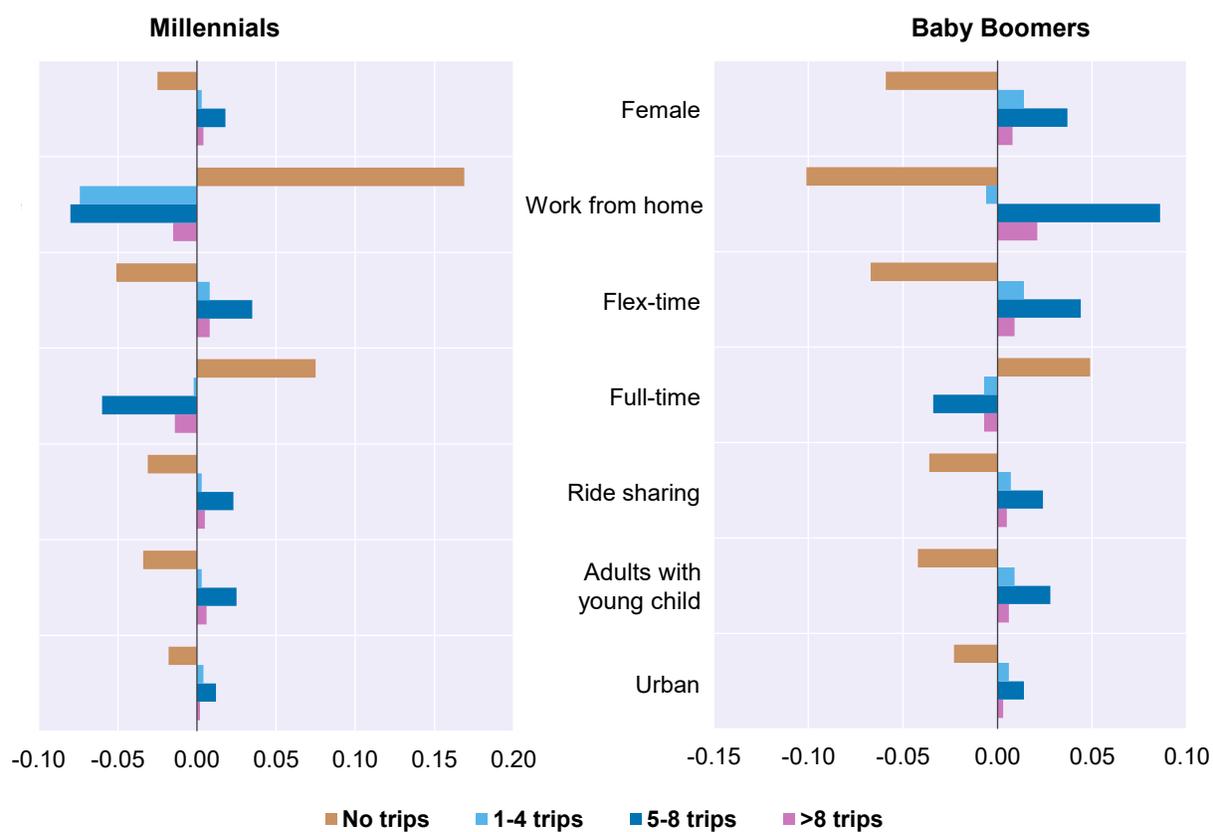

**Figure 2. Comparison of marginal effects of key variables influencing non-work trip frequency among Millennials and Baby Boomers**

**CONCLUSIONS**

This study aimed to analyze and compare the non-work travel behaviors of Millennials and Baby Boomers, focusing on the factors contributing to their differences. Non-work travel, including trips for shopping, recreation, social activities, and errands, is a significant component of overall travel demand. Unlike work-related travel, which is often predictable and concentrated in specific time frames, non-work travel is more flexible, diverse, and occurs throughout the day. To this end, using data from the 2017 National Household Travel Survey, this research employed bootstrapped segmented ordered logit models to capture the variability in non-work travel preferences and trip frequency across these generational groups. By focusing on generational differences, the study provides critical insights into how sociodemographic and technological changes shape travel behavior, with implications for transportation policy and planning.

The results revealed several key distinctions in the contributing factors affecting non-work travel between Millennials and Baby Boomers. For instance, females had a positive association with non-work travel in both cohorts, but female Baby Boomers exhibited a higher likelihood of making such trips. Among Millennials, female heads of household showed a particularly high likelihood of non-work travel. However, the interaction of gender and urban location revealed contrasting trends: female Baby Boomers in urban areas were more likely to travel for non-work purposes, while female Millennials showed reduced likelihood in such settings.

The lifecycle stages further emphasized generational differences. Baby Boomers with young children in the household were significantly more likely to make non-work trips, while Millennials exhibited similar but less pronounced trends. Work-related variables also showed contrasting impacts. Working from home positively influenced non-work travel for Baby Boomers but negatively (albeit not significantly) affected Millennials, suggesting that remote work may reduce Millennials' discretionary travel needs. Flexible work hours, on the other hand, were consistently associated with increased non-work travel across both cohorts. Lastly, ride sharing was positively associated with non-work trips for both groups, with Baby Boomers showing a higher reliance on this mode.

The findings from this study have important implications for transportation planning. The identified differences in non-work travel behavior between Millennials and Baby Boomers underscore the necessity of updating traditional travel demand models. Existing models primarily focus on peak-hour commuting trips, often neglecting the share of discretionary travel. The results suggest that incorporating flexible travel patterns, work from home options, lifecycle stages, and urban-rural dynamics into travel demand models could improve their predictive accuracy and relevance.

Despite these contributions, the study is not without limitations. We only captured the pre-pandemic behavior using the 2017 NHTS data. Future studies could explore the shifts in travel behaviors of Millennials and Baby Boomers during and after the COVID-19 pandemic periods. Additionally, future research should leverage newer and more comprehensive data to examine how non-work travel behaviors of different generations have evolved over time. Also, a deeper investigation into sustainable transportation choices, such as walking and biking, is necessary to understand the generational shifts toward emerging transportation solutions.

## ACKNOWLEDGMENT

The authors would like to acknowledge the support of New York State Department of Transportation (NYSDOT). The opinions, findings, and conclusions in this paper are those of the authors and not necessarily those of NYSDOT.